\RequirePackage{fix-cm}
\documentclass[smallextended]{svjour3}
\smartqed
\usepackage{amsmath,amssymb,paralist,subfigure,graphicx,cite}
\usepackage{algorithmic}
\usepackage{color}

\def\mc{\mathcal}

\newcommand{\ab}[1]{\textcolor{blue}{{#1}}}

\begin{document}
	
	\title{Infection Curve Flattening via Targeted Interventions and Self-Isolation 
	}
	\author{Mohammadreza Doostmohammadian  \and Houman Zarrabi      \and Azam Doustmohammadian \and
		Hamid R. Rabiee
	}
	
	\institute{ M.  Doostmohammadian \at
		Faculty of Mechanical Engineering, Semnan University, Semnan, Iran. \\
		Tel.: +98-23-31533429\\
		\email{doost@semnan.ac.ir}		
		\and		
		H. Zarrabi \at
		Iran Telecom Research Center (ITRC), Tehran, Iran. \\ 
		\email{h.zarrabi@itrc.ac.ir}
		\and
		A. Doustmohammadian \at
		Gastrointestinal and Liver Diseases Research Center, Iran University of Medical Sciences, Tehran, Iran. \\
		\email{mohammadian.az@iums.ac.ir} 
		\and		
		H. R. Rabiee \at
		Department of Computer Engineering, Sharif University of Technology, Tehran, Iran. \\ 
		\email{rabiee@sharif.edu}
		}
	
	\date{Received: date / Accepted: date}

	\maketitle
	
	\begin{abstract}
		Understanding the impact of network clustering and small-world properties on epidemic spread can be crucial in developing effective strategies for managing and controlling infectious diseases. Particularly in this work, we study the impact of these network features on targeted intervention (e.g., self-isolation and quarantine). The targeted individuals for self-isolation are based on centrality measures and node influence metrics. Compared to our previous works on scale-free networks, small-world networks are considered in this paper. Small-world networks resemble real-world social and human networks. In this type of network, most nodes are not directly connected but can be reached through a few intermediaries (known as the small-worldness property). Real social networks, such as friendship networks, also exhibit this small-worldness property, where most people are connected through a relatively small number of intermediaries. We particularly study the epidemic curve flattening by centrality-based interventions/isolation over small-world networks. Our results show that high clustering while having low small-worldness (higher shortest path characteristics) implies flatter infection curves. In reality, a flatter infection curve implies that the number of new cases of a disease is spread out over a longer period of time, rather than a sharp and sudden increase in cases (a peak in epidemic). In turn, this reduces the strain on healthcare resources and helps to relieve the healthcare services.

	\end{abstract}
	
	\section{Introduction} \label{sec_intro}
	Flattening the COVID-19 infection curve is key to ensuring health services aren't overwhelmed by coronavirus cases \cite{nyt}. It implies that the peak number of patients requiring healthcare at a time is reduced. In other words, flatter infection curve means that fewer people will need to be hospitalized at the same time, which can help prevent hospitals from becoming overwhelmed. This is done by both pharmaceutical (e.g., vaccination, medicine) and non-pharmaceutical intervention measures (e.g., social distancing, self-isolation, quarantine). A flattened curve distributes the needs for healthcare over time and keeps the peak of hospitalizations under the healthcare capacity. 
	Recently social network studies have been of interest to investigate how COVID-19 spreads over real human networks \cite{chang2021mobility,karaivanov2020social,block2020social,reyna2021virus} (even cyber-physical contagion of malicious malware over information networks have been studied \cite{brett2021understanding}). These works study the epidemic from the scope of network science and graph theory. In this perspective, the types of the social network model (e.g., scale-free, clustered scale-free, small-world) and its graph-theoretic features (e.g., clustering, small-worldness, assortativity, preferential mixing, community structures) are of importance in the epidemiological network study.

	The literature on the network science and dynamic modelling perspective of the epidemic is quite expansive. Here, we review a few most-relevant works in terms of probabilistic models, graph properties, and network types. It has been shown that, for the susceptible-infected-susceptible (SIS) model, clustering can speed-up propagation of the \textit{co-infected diseases} as compared to non-clustered networks \cite{hebert2015complex}, while, conversely, it slows down the spread of epidemic over hierarchical social networks \cite{grabowski2004epidemic} and raises the epidemic outbreak threshold in single infection outbreak \cite{hebert2010propagation}. In another perspective, \cite{shang2015epidemic} shows that epidemics spread faster over networks with a higher level of overlapping communities. The effect of clustering in social networks is further studied by comparing scale-free and clustered scale-free networks, both flattening the infection curve \cite{SNAM23} and epidemic outbreak in the SIS model \cite{SNAM20}. Clustering plays a key role in the controllability of social networks \cite{me_complex}, Ebola virus transmission \cite{scarpino2015epidemiological}, and respiratory infections epidemic \cite{volz2011effects} among others. Relevant works also show that, under the susceptible-infected-removed (SIR) model, \textit{community lock-downs} are less effective than self-isolation and social distancing \cite{gosak2021community}. 	
	No work in the literature studies how small-worldness affects the infection curve flattening. Few works are focused on the effect of small-worldness on diffusion (of information/disease) process \cite{Nguyen2012UnderstandingAM}, herd immunization \cite{thedchanamoorthy2014influence},
	and epidemic outbreak (by tuning the average path length) \cite{Reppas2011OnTE}. 
	The other relevant works on immunization of epidemic spreading in networks include \cite{wu2015influence,madar2004immunization,wang2020preventing,zuzek2015epidemic,ghalmane2019immunization,li2019suppression}. The work \cite{ghalmane2019immunization}  exploits community structures to control epidemic. Bond percolation models to study immunization are discussed in \cite{madar2004immunization}. Comparison between static and dynamic immunization strategies are discussed in \cite{wu2015influence}. The work \cite{wang2020preventing} proposes novel optimization strategies for community-based immunization of targeted nodes. Mitigation strategies to prevent disease propagation over multi-layer networks is discussed in \cite{zuzek2015epidemic,li2019suppression}. 
	
	In this paper, we study the targeted isolation of individuals in small-world networks modelled based on 
    Watts-Strogatz (WS) model \cite{watts1998smallworld}. Small-world networks are characterized by a high level of clustering, meaning that nodes in the network tend to be highly connected to their immediate neighbours, and short path lengths, meaning that the virus can travel from any node to any other node in the network using a small number of steps. Real social networks also exhibit these characteristics, which is why small-world networks have been used as a model to understand and study social networks.
    In this work, we study if the presence of clusters and high small-worldness in a network is advantageous for targeted interventions to control the spread of an epidemic or not. Identifying and targeting specific individuals (or even clusters or communities) by interventions such as quarantine, self-isolation, contact tracing, and vaccinations can potentially contain the spread of the epidemic and flatten its infection curve over small-world networks (as in any other type of network). The targeted individuals are identified based on centrality measures and isolated to slow down the virus spread. These isolation-based scenarios are shown to be effective in COVID-19 spread both in theory \cite{block2020social,SNAM23} and in reality \cite{nyt}. Our results are not limited to the coronavirus spread, but any other virus-based epidemic can be considered. Flattening the infection curve reduces the burden on healthcare services by giving more time to prepare for the influx of patients and allows for more individualized attention to each patient. 
    
    \ab{The infection curve flattening in this paper is  related to the network immunization problem, although they are two distinct strategies used in the context of managing infectious diseases. They have different goals and methods, but they share some similarities. The aim of this work is to reduce the peak of the infection curve and spread out cases over a more extended period to prevent overwhelming the healthcare systems. The primary objective of infection curve flattening is to slow down the rate of new infections within a population over time. This does not necessarily eliminates the disease entirely but seeks to control its spread by public health measures like social distancing,  quarantine, self-isolation, travel restrictions, and lock-downs. On the other hand, network immunization, also known as targeted immunization, aims to disrupt disease transmission within specific networks or communities by identifying and immunizing key individuals. It focuses on reducing transmission within high-risk clusters and networks by identifying key nodes (individuals or groups) in a transmission network who are most likely to spread the disease. These individuals are then prioritized for vaccination or other preventive measures. The main differences is that infection curve flattening addresses the general population and aims to \textit{slow-down} overall disease spread, whereas network immunization aims to \textit{disrupt} transmission within specific high-risk groups. The similarity is that both strategies aim to reduce the impact of the disease on the population, although through different approaches. Further, both approaches identify and target key individuals (or behaviors) that contribute significantly to disease transmission. Our strategy can be used in conjunction as part of a comprehensive disease control strategy, with infection curve flattening serving as a broader measure to complement network immunization. Interested readers are referred to \cite{SNAM20,liu2021efficient,giakkoupis2005models,pastor2002immunization,thedchanamoorthy2014influence,ghalmane2019immunization,liu2022network} for better understanding of network immunization.}


	\section{Statement of the Problem} 
        \begin{figure}[tbp]
    	\centering
    	\includegraphics[width=2.8in]{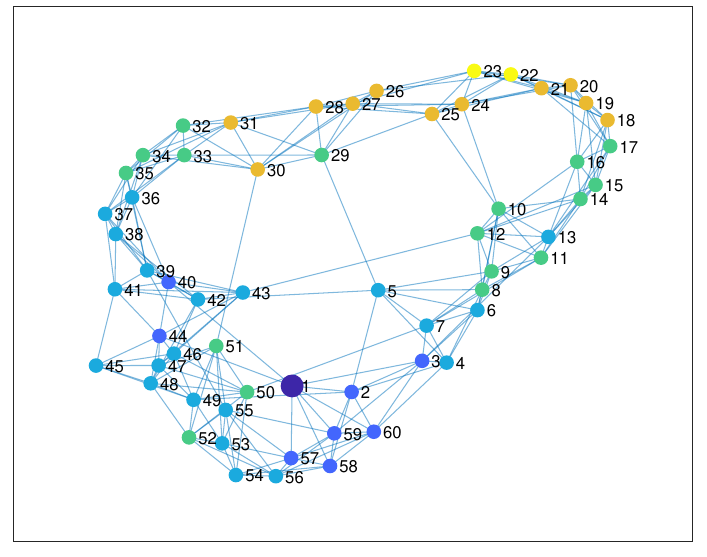}
    	\includegraphics[width=3in]{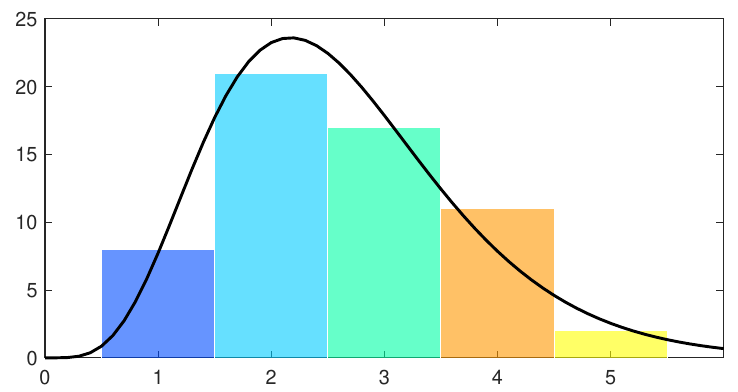}
    	
    	\caption{ An example SW social network and the histogram of its distance distribution from the source node:
    	The nodes represent individuals and the links represent social interactions among people. The large node $1$ is the hypothetically infected node (referred to as the source-node); after one iteration the virus spreads from the infected node to its direct neighbours and at the next iteration from those nodes to their neighbours (2-hop neighbours) and so on. The nodes of the same colour are at the same distance to the source node and, therefore, the virus reaches them after the same number of iterations. The histogram figure below represents the epidemic curve (or the infection curve), where the length of each bar represents the number of infected nodes at the specific distance to the source node (nodes of the same colour). This histogram of the distance distribution (and its fitted curve) makes the so-called infection curve. The idea is to make this epidemic curve flat and long-tailed (by isolating some nodes and hindering the epidemic spread) to prevent a sharp peak of infections in society and to keep the number of cases under the healthcare system's capacity. In the bottom figure, the $x$-axis represents the iterations (and the hop distance) at which the nodes get infected and $y$-axis represents the number of infected nodes.} 
    	\label{fig_sample_hist}
    \end{figure}
    
    This paper studies the spread of infectious diseases over small-world networks. \ab{We consider the infection spread model in \cite{block2020social}. In our model, similar to \cite{block2020social}, we assume that at every iteration all the neighbors of the current infected node get infected. In other words, all the neighboring nodes in contact with the infected individual also get infected at the next iteration/epoch. This model better describes the COVID-19 spread as suggested by \cite{block2020social}. We recall the model  in \cite{block2020social}. In the first iteration, the virus spreads from the seed node to its direct neighbours connected to the seed node. In the second iteration, it spreads to their neighbours, who are at network distance $2$ from the seed node, and so on. Over time, the virus moves along the network ties until all nodes are infected. This implies that the network distance of a node from the infection source (i.e., the seed node) is identical to the number of iterations until the virus reaches it. Then, the distribution of network distances to the source  maps onto the infection curve.} The problem is to find the infection curve of the epidemic and flatten this curve to reach long-tailed distribution associated with the shortest path length. The problem is described as follows and illustrated in Fig.~\ref{fig_sample_hist}. Assume an infected node as the source node of the epidemic spread. At the first iteration, the virus spreads to the neighbours of the infected source node (shown by dark blue in Fig.~\ref{fig_sample_hist}). At the next iteration $2$, the virus spreads from the neighbours to neighbours of the neighbours or 2-hop neighbours (shown by light blue). The shortest path length from the source node to these 2-hop neighbours is 2. Then, at the next iterations, the virus spreads to susceptible 3-hop, 4-hop and 5-hop neighbours (respectively shown by green, orange, and yellow). \ab{To explain more, at every iteration $t$ all the neighboring nodes of the infected nodes get infected and at th next iteration $t+1$ the neighbors of the recently infected nodes (at step $t$) get infected. In this way, the resulting histogram of the infected nodes over different iterations $t+d$ in fact represents the $d$-hop neighbors of the infected source node.} In other words, the $d$-hop neighbours (the nodes at the distance $d$) get infected at the same time after $d$ iterations. Counting the number of $d$-hop neighbours, the histogram represents the distance distribution of the nodes with respect to the infected source node (shown in Fig.~\ref{fig_sample_hist}-bottom). Fitting a Gamma distribution \cite{SNAM23} to this histogram gives the infection curve of the social network. Isolating some of the nodes (or removing some of the links \cite{block2020social}) lengthens the shortest path (distance) from the source node to some other nodes and results in a more flat distance distribution. This implies the idea of infection curve flattening in real human networks to avoid high epidemic peaks and hinder overload to the healthcare system \cite{block2020social}. The problem in this work is how to choose the self-isolated nodes to more flatten the infection curve over SW networks. In this work, the isolated nodes are chosen based on the node influence metrics and centrality measures. However, many graph properties may contribute to the rate of disease spread over the SW networks. In this direction, we study two network structural parameters that affect the infection curve flattening, namely small-worldness and clustering. \textit{To summerize, we change the SW rewiring probability that affects both the clustering and small-worldness of the SW network. Recall that the infection curve is constructed by the distance distribution (i.e., the distribution of shortest path length). We study how shortest path length (and its distribution) changes as we change the rewiring probability (and consequently change the clustering and small-worldness). Therefore, as other network characteristics including the number of linking and average node degrees remain unchanged, one can study how infection curve is affected by these two main network characteristics, clustering and small-worldness.}   
    
    \subsection{Small-World Networks} 
    In a small-world network/graph many nodes are not directly connected, but with a high probability, their neighbours are likely to be connected to each other. This makes it possible to reach most neighbouring nodes with just a few hops or steps from any source node. It is claimed that the shortest path length (or the distance) between two randomly chosen nodes grows proportional to $log(n)$ (with $n$ as the network size). In the perspective of social networks, the small world phenomenon connects strangers through a brief chain of acquaintances and is motivated by \textit{empirical} social networks which show this effect.  
    Small-world networks tend to contain all-to-all connected sub-networks, referred to as \textit{cliques}. In other words, in a clique (almost) any two nodes are adjacent and connected with a link. This further implies a high clustering coefficient in the network. This coefficient can be further tuned by adding triad formations \cite{zaidi2013small} or, similarly, the community structure of the network can be further managed by clique generation \cite{sallaberry2013model}. Moreover, the majority of the nodes can be connected via a short path, indicating a small mean shortest path length within the network. This can be attributed to the abundance of hubs or high-degree nodes, which act as connections/neighbours, bridging the shortest path between other nodes. Network \textit{small-worldness} (or small-worldliness) is quantified by its clustering and path length compared to an equivalent random network (with the same average node degree) and a lattice network. Different definitions for this coefficient are used which are listed in the following:
    \begin{align}
    	\mc{S}_1 = \dfrac{\mc{C}}{\mc{L}}\dfrac{\mc{L}_r}{\mc{C}_r}
    \end{align}
    with $\mc{C},\mc{L}$ and $\mc{C}_r,\mc{L}_r$ as the clustering coefficient and characteristic path length of the graph and its random equivalent \cite{humphries2008network,telesford2011ubiquity}. Based on this metric $\mc{S}_1>1$ (or $\mc{C} \gg \mc{C}_r$  and $\mc{L} \approx \mc{L}_r$) implies that the network resembles a small-world; \textit{however, this metric performs poorly in large-scale}. Another alternative metric is given below:
    \begin{align}
    	\mc{S}_2 = \dfrac{\mc{L}_r}{\mc{L}}-\dfrac{\mc{C}}{\mc{C}_l}
    \end{align}
    with $\mc{C}_l$ as the clustering coefficient of the equivalent lattice and $\mc{L}_r$ as the characteristic path length of the equivalent random graph \cite{telesford2011ubiquity}. Finally, the following metric is the normalized version of $\mc{S}_2$ \cite{neal2017small}, 
    \begin{align}
	\mc{S}_3 = \dfrac{\mc{L}-\mc{L}_l}{\mc{L}_r -\mc{L}_l }\dfrac{\mc{C}-\mc{C}_r}{\mc{C}_l -\mc{C}_r }
    \end{align}     
    with $\mc{L}_l$ as the characteristic path length of the equivalent lattice. 
    
    The most well-known mechanism to build small-world networks is the Watts-Strogatz (WS) model \cite{watts1998smallworld}. This model first builds a regular ring lattice (or cyclic graph) with every node connected to the same number $\frac{k}{2}$ nearest neighbours on each side ($1,2,\dots,\frac{k}{2}$-hop neighbours). Then, the WS model takes every link connecting a node to its $\frac{k}{2}$ rightmost neighbours and randomly rewires it to another node with probability $\beta$ (avoiding self-loops and repetitive links). The lattice-shape structure produces a (locally) highly-clustered network with the random rewiring significantly reducing the average path lengths and network diameter. In fact, for $\beta =0$ the model gives a regular lattice and for $\beta = 1$ it gives an Erdos-Renyi (ER) random network. Clearly, the small-worldness property (i.e., high local clustering and small average path length) are addressed by this model.
    
    \subsection{Clustering}	
    Clustering in a network refers to the tendency of nodes to form clusters or groups of nodes that are more densely connected to each other than to nodes in other parts of the network. It is a measure of the degree to which nodes tend to form cohesive subgroups or communities (e.g. make triads).
    In graph theory, the clustering coefficient is a measure of the proportion of a node's neighbours that are also connected. In real social networks, people tend to form clusters or groups with others who share similar interests, beliefs, or backgrounds. These clusters create highly connected sub-networks within the larger network. 
    
	The common formulation for the (global) clustering coefficient is based on the triplets of nodes, i.e., is defined as the ratio of the closed triplets to all (open and closed) triplets.  Recall that a triplet denotes three nodes connected by either two (open triplet) or three links (closed triplet). From a social network perspective, a triad/triangle implies that the friend of my friend is also my friend. Therefore, high clustering means that two nodes sharing a neighbour are very likely to be connected themselves, i.e., to be neighbors of each other.
	
	The formulation for the clustering coefficient is given below \cite{wasserman1994social}, 
	\begin{align}
		\mc{C} = \dfrac{\mbox{Number of closed triplets}}{\mbox{Number of triplets}} = 3\dfrac{\mbox{Number of triads}}{\mbox{Number of triplets}}
	\end{align}
    where the triad is also referred to as a triangle. For the small-world model, the clustering coefficient is a function of the randomness parameter $\beta$. For the ring lattice $\mc{C}(\beta \rightarrow 0) = \dfrac{3(k-2)}{4(k-1)}$ and for the random ER network the coefficient is $\mc{C}(\beta \rightarrow 1) = \dfrac{k}{n-1}$. In the intermediate range, $\mc{C}$ value remains close to that of the regular network and only falls at relatively high $\beta$.

	\subsection{Shortest Path Length} \label{sec_infect_curve}
\begin{figure}[]
		\centering
		\includegraphics[width=2.5in]{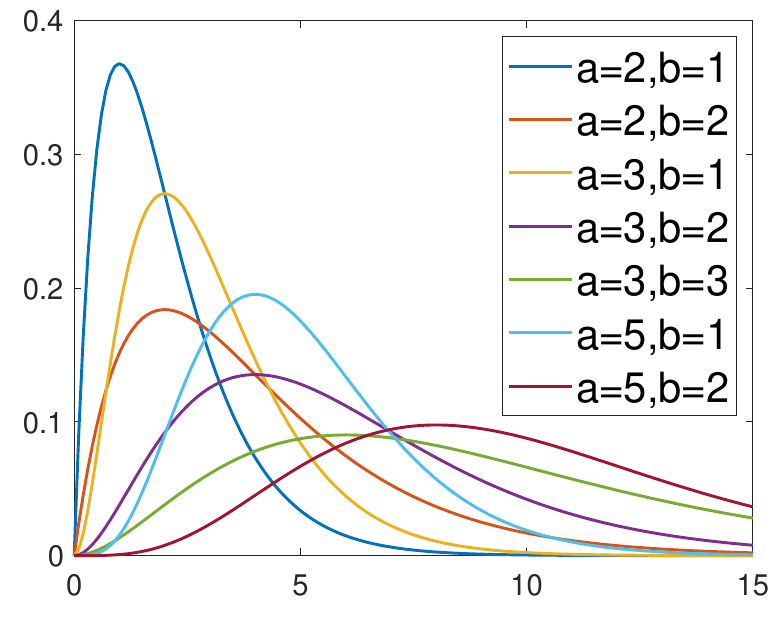}
		\caption{ The PDF of Gamma distributions for different   $a$ and $b$ values as shape and scale parameters. Larger values of scale parameter $b$ imply flatter PDFs.  
		}
		\label{fig_gamma}
	\end{figure}

	In a network, the shortest path (also known as the \textit{distance}) between two nodes is the path with the minimum number of edges that connects them. It is also known as the \textit{geodesic path}. The concept of shortest path is commonly used in graph theory and network analysis and is an important measure of connectivity and accessibility within a network. From a network epidemic perspective, it takes $d$ steps (or iterations) to reach from a node $i$ to another node $j$. In some literature, node $j$ sometimes is referred to as the \textit{$d$-hop neighbour} or  \textit{$d$-hop distant neighbour} of node $i$. For small-world networks, the average path length is a function of $\beta$ and for a regular ring lattice, we have $\mc{L}(\beta \rightarrow 0) \approx \dfrac{n}{2k}$ and scale linearly with $n$. As the network tends toward random network we have $\mc{L}(\beta \rightarrow 1) \approx \dfrac{\ln(n)}{\ln(k)}$. The distance distribution follows a Gamma distribution with its PDF defined as follows:
	\begin{align}
		f(x) = \dfrac{1}{\Gamma(a)b^a}x^{a-1}\exp(-\frac{x}{b})
	\end{align}
	where $a>1$ is the shape parameter and $b>0$ is the scale parameter. Note that, from the figure, larger $b$ values imply flatter Gamma distribution. Some sample PDF Gamma distributions are shown in Fig.~\ref{fig_gamma}.

	\subsection{Node Influence Metrics} \label{sec_centrality}
	
	Centrality measures  are quantitative metrics used to identify the most important or influential nodes within a network. Therefore, they are also referred to as node influence metrics. Some surveys describing different centrality measures over complex and social networks are given in the literature, see \cite{saxena2020centrality,das2018study} for different centralities and their applications. Some well-known examples (particularly related to epidemics) are listed below.
	
	\textbf{Degree:} The node degree refers to the number of edges directly connected to a particular node (i.e., the number of its neighbours) \cite{saxena2020centrality,das2018study}.
	
	\textbf{Betweenness:} This centrality is a measure of a node's importance based on the number of shortest paths that pass through it \cite{freeman1977set}. In other words, it measures how often a node acts as a bridge along the shortest path between other nodes in the network \cite{saxena2020centrality,das2018study}. 
	
	\textbf{Closeness:} This path-based centrality is a measure of a node's importance based on the average distance between that node and all other nodes. This measure calculates the reciprocal of the sum of the distances (shortest paths) to all other nodes \cite{bavelas1950communication}. In other words, it measures how quickly a node can reach all other nodes in the network \cite{saxena2020centrality,das2018study}.
	
	\textbf{Katz:} This degree-based centrality is calculated by summing the number of walks between a node and all other nodes, with the number of walks decreasing exponentially as the length of the walks increases \cite{katz1953new}. The Katz centrality of a node is higher if it is directly connected to other highly central nodes, or if it lies on many paths between such nodes \cite{saxena2020centrality,das2018study}.  
	
	\textbf{Page-Rank:} Similar to Katz score, this is another degree-based centrality quantifying a node's score proportional to the score of the nodes that link to it, as measured by the number and quality of those links \cite{sullivan2007google}. The Page-Rank score tends to be higher for nodes linked to other highly central nodes, and lower for nodes that have few or low-quality incoming links. In simple words, a node $i$ has a high Page-Rank if its neighbours $j$ are highly central, and node $i$ is one of the few neighbours of those nodes $j$ \cite{saxena2020centrality,das2018study}.

	\textbf{Expected Force:} This is another degree-based centrality and entropy-based epidemiological measure. It is a measure of the importance of a node within a network that takes into account both the node's degree and the degree of its neighbours. The expected force centrality of a node $i$ is defined as the sum of the product of the degree of node $i$ and the degree of each of its neighbours divided by the total number of possible edges in the network \cite{lawyer2015understanding}.

	\textbf{Eigenvector:} This centrality (also named Bonacich centrality) denotes the eigenvector associated with the greatest eigenvalue of the adjacency matrix of the network \cite{bonacich1972factoring}. It can be proved that all the entries in this eigenvector are non-negative (this follows from the Perron-Frobenius theorem). To get an absolute score, the eigenvector is sometimes normalized such that the sum over all vertices is $1$ (or $n$). Nodes with high eigenvector centrality are those that are connected to other highly central nodes and thus have a greater ability to influence the behavior of other nodes in the network \cite{das2018study}.
	
	\section{Main Results}
	We study how tuning different network properties flatten the infection curve of the epidemic over the network. Recall that epidemic curve flattening is tightly related to the shortest path length (or, more accurately, the node distance distribution). Based on the preliminaries given in the previous section, the simulations in this section show how self-isolation (or quarantine) affects the epidemic spread over SW networks. More specifically, we isolate the nodes based on the centrality measures and check if network features such as clustering and small-worldness encourages or discourages the infection curve flattening.
	
	First, we compare different network properties for different rewiring probability $\beta$ in WS networks (as the main model for SW networks). We summarized some of these graph properties in Table~\ref{tab_sw} which are averaged over $50$ Monte-Carlo trials of WS networks with  $n=500$ nodes and $k=6$. From the table, larger $\beta$ value implies smaller average shortest path $\mc{L}$, lower clustering $\mc{C}$, higher small-worldness $\mc{S}_2,\mc{S}_3$, and smaller scale parameter $b$. 
	
	\begin{table}[hbpt!]
		\centering
		\caption{How network properties and Gamma parameters change with small-world parameter $\beta$ (averaged over $50$ Monte-Carlo trials).}
		\begin{tabular}{|l|c|c|c|c|c|c|c|}
			\hline
			$\beta$ value & 0.025 & 0.05 & 0.1 & 0.15 & 0.2 & 0.25 & 0.3\\
			\hline
			average shortest path $\mc{L}$ & 6.8  & 5.4 & 4.7 & 4.2 & 4 & 3.9 & 3.9\\
			\hline
			clustering $\mc{C}$ & 0.63 & 0.58 & 0.49 & 0.42& 0.35 & 0.29 & 0.24 \\
			\hline
			small-worldness $\mc{S}_2$ & -0.48 & -0.30 & -0.05 & 0.14 & 0.28 & 0.41 & 0.51\\
			\hline
			small-worldness $\mc{S}_3$ & -0.13  & -0.03 & 0.15 & 0.29 & 0.41 & 0.51 & 0.6\\
			\hline
			scale parameter $b$ & 0.56 & 0.38 & 0.28 & 0.24 & 0.22 & 0.2 & 0.19\\
			\hline
		\end{tabular}
		\label{tab_sw}
	\end{table}

   \subsection{Clustering and Small-Worldness in SW Networks}	
    First, we study clustering. Clusters in a network can both (i) act as natural barriers and hinder the spread of an epidemic and also (ii) facilitate the spread of an epidemic within communities or clusters. The case (i) is because if the connections between the clusters are limited (or sparse), although spreading faster inside the clusters, the virus/disease may struggle to cross over from one cluster to another which reduces the overall epidemic spread (i.e., acting as barriers to cross-cluster transmission). This barrier effect can slow down the epidemic's progression and more flattens the infection curve by limiting its reach to specific clusters or communities. On the other hand, in case (ii), if the interconnections within the clusters are dense and individuals (nodes) densely interact (link) with others within their community, the disease may spread more rapidly within the cluster, leading to a faster and more extensive spread of the epidemic and steeper (high-peak) infection curves. In other words, this can lead to a more rapid spread of the epidemic in the network as a whole, especially if the connections between clusters are also relatively dense.

    Next, we study small-worldness. Small-worldness is a property of social and human networks where most nodes are not directly connected, but can be reached through a small number of intermediate connections. Recall that, it is characterized by high clustering (i.e., individuals tend to have many common neighbours) and short average path lengths (i.e., it is possible to infect most nodes/individuals from any other node/individual in a relatively small number of steps).  Therefore, high small-worldness can facilitate the faster spread of infectious diseases due to the short average path lengths. Infected individuals can rapidly transmit the disease to their close neighbours, who in turn can transmit it to their neighbours, leading to rapid spread within clusters of connected individuals. This can result in a steeper initial rise in the infection curve, as the disease spreads quickly through the network. These factors imply that small-worldness directly affects the spread of infectious diseases, including how the infection curve flattens during an epidemic. Note that high values of small-worldness also imply the formation of clusters or communities of tightly connected individuals (high clustering). If infection occurs within these clusters, it may lead to localized outbreaks, resulting in a spike in the infection curve for those clusters. Once the infection reaches the periphery of the clusters, it may spread through the shortcut links\footnote{The prevalence of the shortcut links is related to the rewiring probability $\beta$.} easily to other parts of the network, which can result in a sharp rise in the overall infection curve.

   \subsection{Self-Isolation Simulation} 
    In order to suppress the spread of infectious disease and reduce the network susceptibility, individuals with high centrality (e.g., network \textit{hubs} or individuals who efficiently bridge the transmission path between different parts of the network) are isolated. If these central nodes become infected, they can transmit the disease to a large number of individuals, potentially resulting in a rapid rise in the infection curve. Therefore, by targeted interventions on these nodes, one can flatten the infection curve as simulated in this section. By changing the $\beta$ value of WS networks, the clustering $\mc{C}$ and small-worldness $\mc{S}_2$ are tuned. We try the simulations for different values of network size $n$ and initial wiring $k$ to show that the results are irrespective of these values. In other words, these show robustness of the results to $n$ and $k$. Also, note that by changing the  $\beta$ value the number of links and average node degrees remain unchanged. Therefore, the network linking is the same and has no effect on the simulation results.

\begin{figure}[]
    	\centering
    	\includegraphics[width=2.25in]{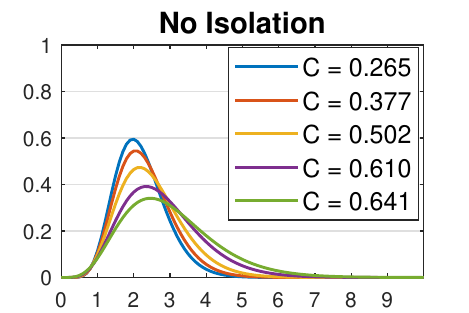}
    	\includegraphics[width=2.25in]{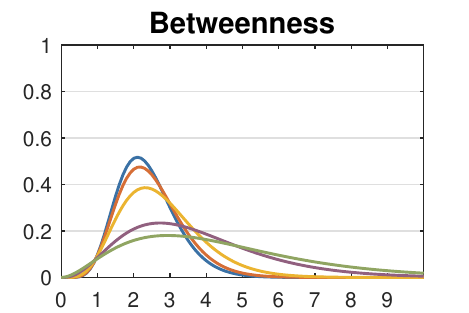}
    	\includegraphics[width=2.25in]{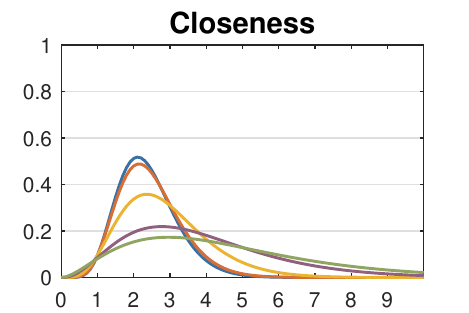}
    	\includegraphics[width=2.25in]{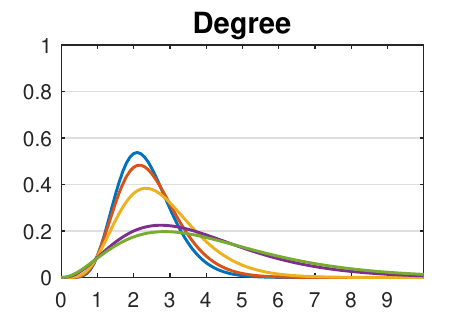} 
    	\includegraphics[width=2.25in]{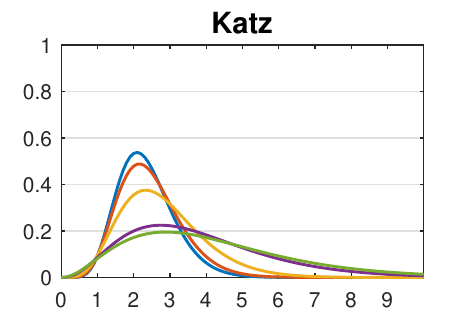}
    	\includegraphics[width=2.25in]{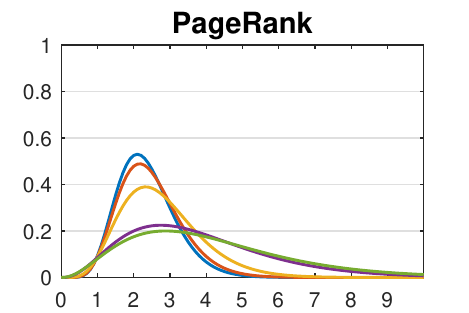}
    	\includegraphics[width=2.25in]{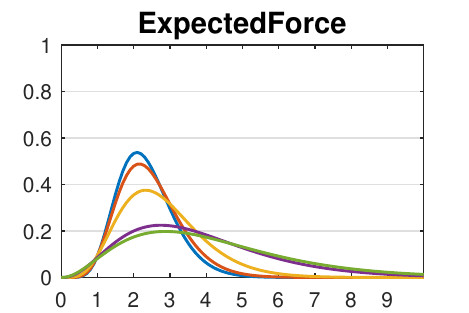}
      	\includegraphics[width=2.25in]{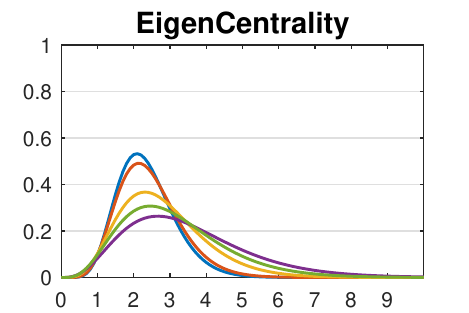}
    	\caption{\ab{The infection curves under targeted interventions and different centrality measures. Some highly central nodes are isolated to flatten the infection curve. These self-isolated nodes are chosen based on their centrality ranks and the results are repeated for different centrality measures. The top-left figure shows infection curve under no isolation and different network clustering. In all the figures, the $x$ axis represents the infection iteration which is the same as the distance to the source node, and the $y$ axis represent the fitted PDF value.} }
    	\label{fig_removal_threshold}
    \end{figure}

    For WS networks, in this paper, Fig.~\ref{fig_removal_threshold} presents the normalized infection curve flattening with $15\%$ self-isolation. The figure on the top-left shows the normalized infection curve under no intervention/isolation while in the rest the infection curves are flattened by $15\%$ node isolation under various centrality measures and node influence metrics. Recall that the curves are normalized by the total number of shortest paths. The associated clustering coefficient values are given in the first figure. The simulations are averaged over $100$ Monte-Carlo (MC) trials. The clustering coefficient is tuned by changing the rewiring probability $\beta = 0.025,0.05,0.1,0.2,0.3$ while the number of linking, network connectivity, and average node degrees remain the same ($k=6$ and $n=100$ for all WS networks). This gives the clustering coefficients as $\mc{C} = 0.660,0.634,0.540,0.423,0.337$. It is clear from the figure that increasing the clustering coefficient dramatically flattens the infection curve. This implies that, following the case (i), the connections between the clusters are limited (or sparse), as the number of linking in the WS model is fixed, by decreasing the $\beta$ value (and increasing $\mc{C}$) there are fewer options for the cross-cluster transmission of the disease, and in turn, flattens the infection curve. 
    Moreover, in terms of comparing the centralities,  self-isolation via path-based centralities, e.g. closeness, more flattens the infection curves.  

       Other than the qualitative comparison in the figures, we performed quantitative comparisons here. The peak values of the distance distribution associated with~Fig.~\ref{fig_removal_threshold} are given in Table~\ref{tab_sw_peak}. These peak values represent the max number of individuals infected simultaneously at the same iteration of the virus spreading over the network (see Fig.~\ref{fig_sample_hist} for more illustration). Recall that the fitted curves in Fig.~\ref{fig_removal_threshold} are normalized by the total number of shortest-paths. Both clustering coefficient $\mc{C}$ and small-worldness $\mc{S}_2$ are given for better comparison. As it is clear from the table and the figures, self-isolation via path-based centralities (i.e., betweenness and closeness) results in lower peak values of the distance distribution and, thus, the fitted infection curves are flatter.

   \begin{table}[hbpt!]
	\centering
	\caption{The peak values of the distance distributions associated with Fig.~\ref{fig_removal_threshold} before curve normalization \ab{($100$ MC trials, $n=100$, $k=6$, $15\%$ self-isolation)}. Recall that the infection curves are in fact the normalized PDFs fitted to these distance distributions. The small-worldness and clustering values are also given for comparison.}
	\ab{\begin{tabular}{|c|c|c|c|c|c|c|c|c|c|c|}
			\hline
			$\beta$ & $\mc{S}_2$  & $\mc{C}$  & None  & Bet  & Close & Deg & Katz & Page & Exf & Eig\\
			\hline
			0.3	& 0.481 & 0.265 & 4013  & 3047 & 3037 & 3115 & 3122  & 3118 & 3118 & 3108 \\
			\hline
			0.2 & 0.302  & 0.377 & 3822 & 2902 & 2892 & 2932 & 2996 & 2986 & 2986 & 3016 \\
			\hline
			0.1 & 0.052  & 0.502 & 3299 & 2278 & 2158 & 2282 & 2275  & 2332  & 2226 & 2272 \\
			\hline
			0.05 & -0.117 & 0.610 & 2276 & 1369 & 1227 & 1378 & 1378 & 1378 & 1378 & 1648 \\
			\hline
			0.025& -0.275 & 0.641 & 2104 & 923 & 870 & 1066 & 1074 & 1084 & 1070 & 1844 \\
			\hline		
	\end{tabular}}
	\label{tab_sw_peak}
\end{table}	
    
\begin{figure}[]
   	\centering
   	\includegraphics[width=2.25in]{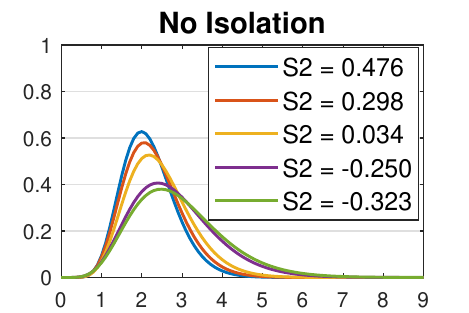}
   	\includegraphics[width=2.25in]{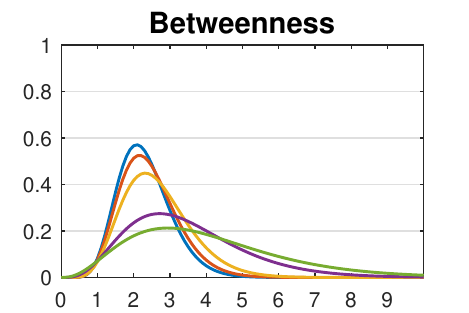}
   	\includegraphics[width=2.25in]{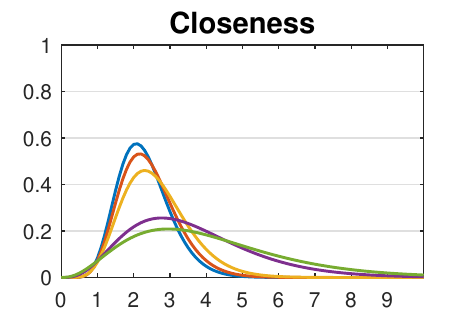}
   	\includegraphics[width=2.25in]{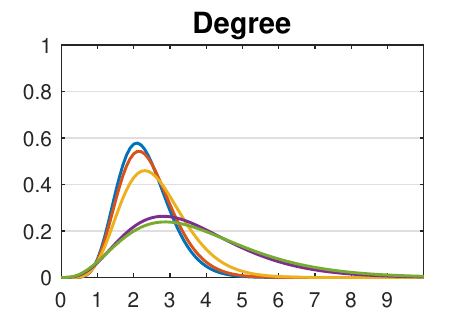} 
   	\includegraphics[width=2.25in]{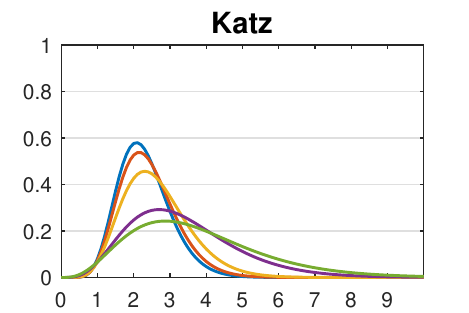}
   	\includegraphics[width=2.25in]{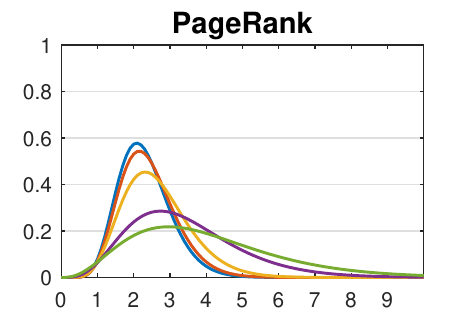}
   	\includegraphics[width=2.25in]{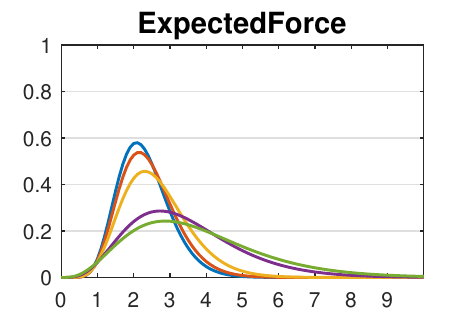}
   	\includegraphics[width=2.25in]{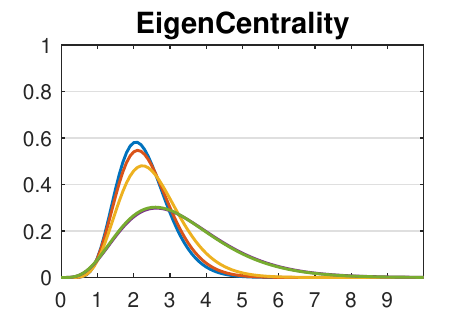}
   	\caption{\ab{The infection curves under targeted interventions and different centrality measures similar to Fig.~\ref{fig_removal_threshold}. The isolated target nodes are chosen based on different centrality measures to see how quarantine of highly central nodes flattens the infection curve. The top-left figure shows the infection curve under no isolation and different network small-worldness. In all the figures, the $x$ axis represents the infection iteration which is the same as the distance to the source node, and the $y$ axis represent the fitted PDF value.}} 
   	\label{fig_removal2}
   \end{figure}

    Fig.~\ref{fig_removal2} presents the infection curve flattening with $14\%$ self-isolation. The top-left figure is with no isolation (i.e., no intervention) and in the rest of the figures, $14\%$ of the nodes are isolated based on different centrality ranks, which significantly flattens the infection curve. The illustrated simulations are averaged over $100$ MC trials, where the small-worldness in the WS model changes by the rewiring probability $\beta = 0.025,0.05,0.1,0.2,0.3$. Recall that, following the WS model (with $k=8$ and $n=150$ in this simulation), the number of linking, network connectivity, and average node degrees remain the same. We compared the $\mc{S}_2$ small-worldness measure\footnote{Regarding the other two small-worldness measures, $\mc{S}_1$  is not properly defined in large-scale and $\mc{S}_3$ changes similar to the $\mc{S}_2$ measure and, therefore, are skipped here.}, clustering, and peak values under different centrality-based isolation for this example.  It is clear from the figure that decreasing the small-worldness dramatically flattens the infection curve. Further, note that isolation based on closeness centrality more flattens the infection curve as compared to other centrality measures. 
   The peak values of the distance distribution associated with Fig.~\ref{fig_removal2} are given in Table~\ref{tab_sw2_peak} where the values are not normalized. Both clustering coefficient $\mc{C}$ and small-worldness $\mc{S}_2$ are given for more comparison. Similarly, this figure and table also show that self-isolation via path-based betweenness and closeness centralities results in lower peak values and, thus, better flattening of the infection curves. 
   
   \begin{table}[hbpt!]
	\centering
	\caption{The peak values of the distance distributions (before curve normalization), small-worldness and clustering values associated with Fig.~\ref{fig_removal2} \ab{($100$ MC trials, $n=150$, $k=8$, $14\%$ self-isolation)}.}
	\ab{\begin{tabular}{|c|c|c|c|c|c|c|c|c|c|c|}
		\hline
		$\beta$ & $\mc{S}_2$  & $\mc{C}$  & None  & Bet  & Close & Deg & Katz & Page & Exf & Eig\\
		\hline
	    0.3	& 0.476 & 0.292 & 7801  & 7364 & 7328 & 7411 & 7350 & 7401 & 7404 & 7384 \\
		\hline
		0.2 & 0.298  & 0.382 & 7595 & 7187 & 7148 &  7360 & 7284  & 7354 & 7341 & 7307 \\
		\hline
		0.1 & 0.034  & 0.516 & 7064 & 6391 & 6326 & 6596 & 6524  & 6473 & 6525 & 6858 \\
		\hline
		0.05 & -0.250 & 0.630 & 6066 & 4125 & 3506 & 3812 & 4415  & 4392 & 4308 & 4424 \\
        \hline
        0.025& -0.323 & 0.658 & 5815 & 3033 & 2803 & 3652 & 3661  & 3280 & 3548 & 4405 \\
        \hline		
	\end{tabular}}
	\label{tab_sw2_peak}
\end{table}		

	\section{Discussions and Conclusions}
	This paper investigates how different network features such as clustering and small-worldness affect the centrality-based epidemic curve flattening over SW networks. We summarize our results as follows:
	
	\begin{enumerate}[(i)]
		\item There are two types of centrality affecting the infection curve flattening: (i) degree-based centrality can be useful for identifying highly connected individuals or "hubs" who may have a greater potential to spread the infection. Targeting these individuals for interventions, such as quarantine or vaccination, can help mitigate the spread of the disease and potentially flatten the infection curve. (ii) Measures such as betweenness or closeness centrality capture the extent to which a node lies on the shortest paths between other nodes. In the context of infectious diseases, path-based centrality can be beneficial in identifying individuals who act as bridges or bottlenecks in the spread of the infection. By targeting these individuals for interventions it may be possible to disrupt the transmission pathways and flatten the infection curve.
		Our results show that, for our single infection-source iterative transmission, node isolation via path-based centralities (e.g. closeness and betweenness) are more effective on flattening the infection curve over SW networks. For more illustration on this, check the quantitative comparisons for different centralities given in Table~\ref{tab_sw_peak} and \ref{tab_sw2_peak} and qualitative comparisons in Fig.~\ref{fig_removal_threshold} and \ref{fig_removal2}. In general, the peak values for path-based centralities in Table~\ref{tab_sw_peak} and \ref{tab_sw2_peak} are lower, implying flatter infection curves, while for degree-based centralities the peak values are higher, implying taller and narrower infection curves. This conclusion confirms our previous results over scale-free (SF) networks \cite{SNAM23} and targeted node control to derive the SF networks towards the healthy-state (SIS model) \cite{SNAM20}. Similar results are claimed by \cite{thedchanamoorthy2014influence}, saying that betweenness-based immunization is the best strategy in static networks. However, it might be the case that in other types of networks and other compartmental models of epidemic different centrality-based isolation work better. 
		\item Our results show that increasing the clustering in the SW networks by decreasing the rewiring probability $\beta$  flattens the infection curve more effectively. 
		network clustering can affect the spread of the epidemic in complex ways (depending on the specific characteristics of the network), it can either slow down the spread of the disease through localized outbreaks and barriers to cross-cluster transmission or enhance the spread within clusters. However, high clustering likely slows down the spread of the disease in SW networks. While the effectiveness of \textit{targeted} interventions can also influence the impact of clustering on the infection curve (see \cite{SNAM23} for SF networks). The network clustering may facilitate the spread of the disease in other types of networks, influence the resilience of the network, and guide targeted interventions for epidemic control (see \cite{SNAM20}) in various network types.
		\item In SW networks with high small-worldness the presence of more shortcuts in the network can facilitate the spread of the infection to more distant parts of the network, and the overall infection spreads more quickly than in random and regular networks. This implies that SW networks with higher small-worldness parameter have steeper (narrower) infection curves. Such high-peak infection curves imply more burden on the healthcare systems.   
	\end{enumerate}
	
	As a future direction, one can further consider link isolation (or link removal) which models social distancing among individuals. The targeted links can be chosen based on link-centralities, e.g., shortest path-based, to flatten the infection curve. The results can be extended to the analysis of epidemic growth rates over synthetic and cyber-networks \cite{csl2020}. Another future direction is to consider that a portion (less than $1$) of $d$-hop neighbors get infected at every iteration. This makes the infection curve different from the deterministic distance distribution. For this case an infection probability can be assigned to every link and the \textit{probabilistic} distance distribution is modeled to find the infection curve. Some existing works discuss such distribution models over Erdos-Renyi networks with different link probabilities, see  \cite{PhysRevE.76.066101} for example.
	
	\section*{Acknowledgements}
	 The authors acknowledge the use of some MATLAB codes from Koblenz Network Collection (KONECT) \cite{konect}.

	\bibliographystyle{spmpsci} 
	\bibliography{bibliography}
	
\end{document}